\documentclass[aps,pre,reprint,amsmath,amssymb,showpacs,floatfix,longbibliography]{revtex4-1}
\usepackage{graphicx}
\usepackage{bm}
\usepackage{color}
\usepackage{lipsum}
\usepackage{soul}

\def\la{\left\langle}
\def\ra{\right\rangle}

\def\Fe{f}
\def\Febf{\bm{f}}
\def\T{\mathcal{T}}
\def\parabf{; w(\bm{x},\bm{v})}

\begin{document}

\title{Free diffusion bounds the precision of currents in underdamped dynamics}

\author{Lukas P. Fischer}
\author{Hyun-Myung Chun}
\altaffiliation[New Adress: ]{Department of Biophysics, University of Michigan, 
Ann Arbor, Michigan, 48109, USA}
\author{Udo Seifert}
\affiliation{II. Institut f{\"u}r Theoretische Physik, Universit{\"a}t Stuttgart,
70550 Stuttgart, Germany}

\date{\today}

\begin{abstract}
The putative generalization of the thermodynamic uncertainty relation (TUR) to underdamped dynamics is still an open problem. So far, bounds that have been derived for such a dynamics are not particularly transparent and they do not converge to the known TUR in the overdamped limit. Furthermore, it was found that there are restrictions for a TUR to hold such as the absence of a magnetic field. In this article we first analyze the properties of driven free diffusion in the underdamped regime and show that it inherently violates the overdamped TUR for finite times. Based on numerical evidence, we then conjecture a bound for one-dimensional driven diffusion in a potential which is based on the result for free diffusion. This bound converges to the known overdamped TUR in the corresponding limit. Moreover, the conjectured bound holds for observables that involve higher powers of the velocity as long as the observable is odd under time-reversal. Finally, we address the applicability of this bound to underdamped dynamics in higher dimensions.
\end{abstract}
\maketitle

\section{Introduction}
A recent result in stochastic thermodynamics~\cite{Seifert:2012es} is the thermodynamic uncertainty relation (TUR)~\cite{Barato:2015kq,Gingrich:2016ip,horo19}. It bounds the precision of any current by the mean rate of entropy production $\sigma$ of the system in the steady state.
This bound embodies the intuitive picture that increasing the precision of currents comes with the cost of higher dissipation.
In detail, the TUR states
\begin{equation}\label{eq:overdamped_TUR}
    \mathcal{Q}(\T) 
    \equiv \epsilon(\T)^{2} \sigma \mathcal{T}
    \geq 2 k_{\rm B}
\end{equation}
for the uncertainty $\epsilon(\T)$ of a time-integrated current $Y(\T)$ measured up to time $\T$. The uncertainty (the inverse precision) $\epsilon(\T)$ is defined as
\begin{equation} \label{eq:defPrecision}
	\epsilon(\T)^2 \equiv \frac{\mathrm{Var}[Y(\T)]}{\langle Y(\T)\rangle^2}
\end{equation}
where $\langle \cdot \rangle$ denotes the ensemble average taken over the steady state and ${\rm Var}[Y]$ is the variance of $Y$. In the following, we will refer to the quantity $\mathcal{Q}(\T)$ as the \emph{uncertainty product}.

The TUR was first proven for continuous-time Markovian dynamics in the long-time limit for discrete states~\cite{Gingrich:2016ip} and for overdamped continuous states~\cite{Polettini:2016hu,Gingrich:2017jm} exploiting the level 2.5 large deviation functional~\cite{Maes:2008kv,Maes:2008bu,Barato:2015hh,Hoppenau:2016cu}.
Later, the validity of the TUR was extended to finite times $\T$~\cite{Pietzonka:2017iq,Horowitz:2017ut}, as presented in Eq.~\eqref{eq:overdamped_TUR}.

The significance of the TUR is not only of a theoretical nature. It also has practical implications from a more applied point of view. For example, the TUR has been used to derive an upper bound on the efficiency of molecular motors~\cite{Pietzonka:2016ih} and of heat engines~\cite{Pietzonka:2018de}.
While the original TUR in \eqref{eq:overdamped_TUR} holds for systems in a steady state, modified versions of the TUR for systems with a transient initial condition \cite{Dechant:2018ga,Dechant:2018ea} and periodically driven systems~\cite{Proesmans:2017ic,Barato:2018hb,Koyuk:2019kr,Koyuk:2019dm} have also been derived. Furthermore, quantum versions of the TUR have been considered (see e.g.~\cite{kuma18,Brandner:2018cr,ptas18,carr19}).

Attempts to generalize the TUR to other, non-current observables showed that the choice of the observable $Y$ is crucial. For overdamped Langevin systems described by the degrees of freedom $\bm{x}(t)$, the TUR holds for time-integrated currents of the form 
\begin{equation}\label{eq:overdamped_current}
	Y^{(\circ)}\left(\T; \bm{w}(\bm{x})\right) =  \int_{t=0}^\T \bm{w}(\bm{x}(t))\circ d\bm{x}(t)
\end{equation}
with an arbitrary weight function $\bm{w}(\bm{x})$. Here, $\circ$ denotes an integration in the Stratonovich sense.
As appropriate for a current, the sign of $Y$ is reversed under time-reversal.
For observables that are symmetric under time-reversal, the relation \eqref{eq:overdamped_TUR} does not hold. Instead, the precision is bounded by another quantity that measures the activity of the dynamics~\cite{garrahan_simple_2017,DiTerlizzi:2018kz}.

The validity of the TUR for systems with degrees of freedom that are odd under time reversal still remains an open problem. Prominently, such systems include underdamped Langevin systems in which the velocity must be reversed under time reversal. Although bounds have been derived for such a dynamics~\cite{Fischer:2018fi,Dechant:2018vu,VanVu:2019uu,Lee:2019uy}, they are not as transparent and often not as tight~\cite{Lee:2019uy} as the original, overdamped TUR, Eq.~\eqref{eq:overdamped_TUR}.  Moreover, recent results showed that the original TUR does not hold in general for underdamped dynamics. First, it has been argued that the weight function $\bm{w}$ in the time-integrated observable $Y$ can depend on all degrees of freedom, in particular on the velocity~\cite{VanVu:2019uu}. In this case the weight can be chosen in a way that renders the observable $Y$ even under time-reversal. As for overdamped motion, the uncertainty product can become smaller than the original TUR for such time-symmetric observables. Second, generalized forces that depend not only on the position but also on the velocity can lead to a smaller uncertainty product~\citep{Chun:2019hm,Lee:2019uy}. One example for such generalized forces is the Lorentz force induced by a magnetic field. Third, for finite times the uncertainty product falls below the overdamped TUR~\cite{VanVu:2019uu}. 

In spite of the fact that the TUR has not been proven for general underdamped dynamics, no counterexamples are known in the long-time limit~\cite{Fischer:2018fi} as long as the acting forces do not depend on the velocity~\cite{Chun:2019hm}. Its validity in this limit has been proven only in the linear response regime in absence of a magnetic field that breaks the reversibility of the dynamics~\cite{Brandner:2018cr,Macieszczak:2018jv}. Even though the TUR has not been proven for underdamped dynamics, we call it a ``violation of the TUR'' if Eq.~\eqref{eq:overdamped_TUR} does not hold.

In this paper, we will address the issue of an underdamped TUR in more depth. First, we show that the finite-time behavior of the underdamped uncertainty product $\mathcal{Q}$ is inherently different compared to the overdamped case. In more detail, $\mathcal{Q}$ is generally linear for small times and thus violates the overdamped TUR. We then consider free diffusion with drift for observables that are both even and odd under time reversal. Whereas the uncertainty product of even observables becomes zero in the linear response limit, odd observables show a finite minimum. By extensive numerical simulations, we conjecture that the uncertainty product for odd observables in driven underdamped diffusion across a one-dimensional potential is bounded from below by the corresponding value of free diffusion. This free diffusion bound (FDB) captures the short-time behavior in the ballistic regime. For large times or in the overdamped limit, the overdamped finite-time TUR can be recovered. Finally, we consider two-dimensional systems to assess the validity of the conjecture for multi-dimensional Langevin dynamics.


\section{Underdamped dynamics\label{sec:underdampedDynamics}}
\subsection{Equations of motion}
The dynamics of an underdamped particle of mass $m$ with friction coefficient $\gamma$ and constant temperature $T$ is described by the Langevin equation
\begin{align}\label{eq:underdampedLangevin}
	\dot{\bm{x}}(t)  &= \bm{v}(t), \nonumber\\ 
	m\dot{\bm{v}}(t) &= \bm{F}(\bm{x}(t)) - \gamma\bm{v}(t) + \bm{\xi}(t)
\end{align}
where $\bm{x}$ is the position of the particle, $\bm{v}$ is its velocity, and $\bm{F}(\bm{x}) \equiv -\nabla_x V(\bm{x}) + \Febf $ is the force that stems from a conservative potential $V$ and some external driving $\Febf$. Finally, $\bm{\xi}$ denotes the fluctuating force characterized by independent, zero mean Gaussian  white noise with correlations $\langle \xi_i(t) \xi_j(t')\rangle = 2T\gamma\delta(t-t')\delta_{ij}$ where $i$ and $j$ are Cartesian indices. Here and in the following, we set the Boltzmann constant to unity, $k_\text{B} = 1$. We further assume that the system relaxes to a (non-equilibrium) steady state in the long-time limit. 

Along the stochastic trajectories of the particle, the first law of (stochastic) thermodynamics represents energy conservation. The rate of change of the internal energy $U$ of the particle is given by \cite{seki98}
\begin{equation} \label{eq:energy_conservation}
	\dot{U}(t)
	= m\bm{v}(t) \circ \dot{\bm{v}}(t) 
	+ \nabla_{x} V(\bm{x}(t)) \cdot \bm{v}(t) = \dot{Q}(t) + \dot{W}(t)
\end{equation}
where we have identified the rate of heat exchanged with the medium as
\begin{equation} \label{eq:heat_reat}
	\dot{Q}(t) = \bm{v}(t) \circ \left( -\gamma \bm{v}(t) + \bm{\xi}(t) \right)
\end{equation}
and the rate of work done on the particle as 
\begin{equation}  \label{eq:work_rate}
	\dot{W}(t) = \Febf \cdot \bm{v}(t) .
\end{equation}

When driven out of equilibrium by a finite force $\Febf$, the system permanently dissipates heat into the surrounding medium through friction. This non-equilibrium character is captured by the mean entropy production rate
\begin{equation}\label{eq:entropyProduction}
	\sigma = -\frac{1}{T}\langle \dot{Q} \rangle = \frac{1}{T}\langle \Febf \cdot \bm{v}\rangle .
\end{equation}
Here, the last equality follows from equation \eqref{eq:energy_conservation} where we use that in the steady state the internal energy becomes constant on average.

Extending the notations from the original, overdamped TUR, we consider time-integrated observables of the form
\begin{equation}
	Y^{(\circ)}(\T; \tilde{\bm{w}}(\bm{x}, \bm{v})) = \int_{t=0}^\T \tilde{\bm{w}}(\bm{x}(t), \bm{v}(t))\circ d\bm{x}(t)
\end{equation}
where we can replace the Stratonovich integration $\circ\,d\bm{x}(t)$ with the regular Riemann integration $\bm{v}(t)dt$. As a result, the observable $Y^{(\circ)}$ becomes the regular integral
\begin{equation}\label{eq:underdampedObservables}
	Y(\T\parabf) \equiv \int_0^\T w(\bm{x}(t), \bm{v}(t)) dt
\end{equation}
along the trajectory $(\bm{x}(t), \bm{v}(t))$ with weight 
\begin{equation}
	w(\bm{x}, \bm{v}) \equiv \tilde{\bm{w}}(\bm{x}, \bm{v}) \cdot \bm{v}. 
\end{equation}
In analogy to the overdamped results, we for now allow the weight function to depend on all degrees of freedom, especially the velocity. An example for such an observable is the integrated work up to time $\T$, Eq.~\eqref{eq:work_rate}, with $w(\bm{x},\bm{v}) = \Febf\cdot\bm{v}$.

For such an observable, Eq.~\eqref{eq:underdampedObservables}, the uncertainty product
\begin{equation}\label{eq:uncertaintyProduct}
	\mathcal{Q}(\T\parabf) \equiv \frac{\mathrm{Var}[Y(\T\parabf)]}{\la Y(\T\parabf) \ra^2}\sigma \T
\end{equation}
involves the first and second moment of $Y$. It is worth noting that a constant factor in the weight does not change the uncertainty product.

While the first moment can be calculated from the steady state distribution directly, calculating the second moment is more involved. Its time evolution follows the ordinary differential equation
\begin{equation}\label{eq:ode_YY_gen}
	\frac{d}{d\T}\langle Y^2\rangle = 2\la Y \, w\ra .
\end{equation}
Here, the arguments of $Y$ and $w$ are dropped for better readability. The time evolution of the average on the right hand side can, in turn, be calculated using It\^o's Lemma. After inserting the Langevin equation \eqref{eq:underdampedLangevin} and using that ensemble averages containing the fluctuating force in first order vanish, we arrive at
\begin{align}\label{eq:ode_Ywv_gen}
	\frac{d}{d\T}\langle Y \,w&\rangle = \la w^2\ra + \la Y \, \left(\nabla_x w\right) \bm{v}\ra \\
	& +\frac{1}{m}\la Y \, \left(\nabla_v w\right)\left( \bm{F}(\bm{x}) - \gamma\bm{v}\right)\ra + \frac{T\gamma}{m^2} \la Y \, \Delta_v w \ra  \nonumber
\end{align}
where the gradient with respect to $\bm{x}$ and $\bm{v}$, respectively, is denoted as $\nabla_{x,v}$, and $\Delta_v \equiv \sum_i \partial^2 / \partial v_i^2$ is the Laplace-operator with respect to the velocity.

\subsection{Short-time behavior}

The inertia of underdamped motion has significant impact on the uncertainty product $\mathcal{Q}$ of an observable. In contrast to overdamped motion, it introduces a non-linear time-dependence in the variance of $Y$. This can be shown by taking the time derivative on both sides of Eq.~\eqref{eq:ode_YY_gen} and, subsequently, plugging in Eq.~\eqref{eq:ode_Ywv_gen}. Since $Y$ vanishes for $\T = 0$ by definition, all ensemble averages involving $Y$ vanish as well. Consequently, the variance simplifies to
\begin{equation}\label{eq:VarY_smallTime}
	\mathrm{Var}\left[ Y(\T\parabf) \right] = \mathrm{Var}\left[w\right] \T^2 + \mathcal{O}(\T^3).
\end{equation}
This quadratic dependence is a result of the deterministic equation of motion for $\bm{x}$ which results in a ballistic regime for short times. For longer times, the effect of the noise enters and the velocity decorrelates thus giving rise to the expected linear behavior of the variance.

The ballistic regime in the variance of $Y$ also changes the characteristics of the uncertainty $\epsilon$, Eq.~\eqref{eq:defPrecision}. While in the overdamped limit the uncertainty scales with $\T^{-1}$ for all times, for underdamped dynamics it is of order one in the ballistic regime. As a result, the uncertainty product generally becomes linear
\begin{equation}\label{eq:uncertaintyProduct_ballistic}
	\mathcal{Q}(\T\parabf) = \frac{\mathrm{Var}[w]}{\la w \ra^2}\sigma \T + \mathcal{O}(\T^2)
\end{equation}
thus violating the overdamped TUR and even approaching $0$ in the limit $\T\rightarrow 0$. The reported violations of the overdamped TUR for the (odd) particle current in Ref.~\cite{VanVu:2019uu} can be attributed to this effect.

\subsection{Relation to a bound based on the detailed fluctuation theorem}
On first sight, the linear order of $\mathcal{Q}$ in time seems to contradict a proof of the overdamped TUR for small times that is solely based on the detailed fluctuation theorem for entropy production~\cite{Pietzonka:2017iq}. This proof can, however, not be generalized to underdamped motion as the total entropy production does not follow a detailed fluctuation theorem in a non-equilibrium steady state (NESS).

In underdamped dynamics, the entropy production of a certain trajectory $\Gamma_\T = \lbrace (\bm{x}(t), \bm{v}(t)) | t \in [0, \T] \rbrace$ can be written as
\begin{equation}\label{eq:entropyProduction_traj}
	\Delta S[\Gamma_\T] = \ln \frac{\mathcal{P}[\Gamma_\T]}{\mathcal{P}^\dagger [\Gamma_\T^\dagger ]}
\end{equation}
with the time-reversed trajectory $\Gamma_\T^\dagger = \lbrace (\bm{x}(\T-t), -\bm{v}(\T-t)) | t \in [0, \T] \rbrace$. As usual $\mathcal{P}$ and $\mathcal{P}^\dagger$ denote the path weights associated with a certain trajectory under the original dynamics and its time-reversed complement, respectively. To get the correct stochastic contribution to the entropy production, the complement path weight $\mathcal{P}^\dagger$ must be chosen as \cite{spin12}
\begin{equation}
	\mathcal{P}^\dagger [\Gamma_\T^\dagger ] \equiv \mathcal{P}[\Gamma_\T^\dagger | (\bm{x}(\T), -\bm{v}(\T))]\;p(\bm{x}(\T), \bm{v}(\T)).
\end{equation}
In particular, the initial probability of the complementary weight is given by the final distribution $p(x(\T), v(\T))$ of the original process. Since $\mathcal{P}^\dagger\neq\mathcal{P}$ due to the different initial probabilities, a detailed fluctuation theorem does not follow directly from the definition of the entropy production, Eq. \eqref{eq:entropyProduction_traj}, in contrast to the overdamped case.

In principle, it is possible to follow the proof of the short-time version of the TUR in~\cite{Pietzonka:2017iq} with another irreversibility measure given by
\begin{equation}\label{eq:irreversibility_traj}
	\Delta \Psi[\Gamma_T] \equiv \ln \frac{\mathcal{P}[\Gamma_\T]}{\mathcal{P}[\Gamma_\T^\dagger ]} .
\end{equation}
This functional involves the time-reversed trajectory $\Gamma_\T^\dagger$ and a complementary path weight given by $\mathcal{P}$ as for the forward process. From this definition, a detailed fluctuation theorem follows directly. The functional $\Delta\Psi$ coincides with the entropy production in the long-time limit, where the boundary terms become irrelevant. In the short-time limit, however, this irreversibilty measure converges to the finite value 
\begin{equation}\label{eq:irreversibility_smallTimes}
	\lim_{\T\rightarrow 0} \Delta \Psi[\Gamma_\T] = \ln\frac{p(\bm{x}(0), \bm{v}(0))}{p(\bm{x}(0), -\bm{v}(0))}
\end{equation}
that involves only the initial distribution. Iterating the same steps as described in Ref.~\cite{Pietzonka:2017iq} results on the following bound on the precision
\begin{equation}\label{eq:irreversibility_bound}
	\frac{\mathrm{Var}[ Y(\T)]}{\la Y(\T) \ra^2} \geq \frac{2 - \la \Delta\Psi(\T)\ra}{\la \Delta\Psi(\T)\ra}.
\end{equation}
This bound is valid for non-equilibrium steady states and for all times $\T$.

Unfortunately, the result is trivial in both the long-time and the short-time limit. As heat is constantly dissipated in the medium in a NESS, the irreversibility measure $\la \Delta\Psi(\T)\ra$ grows with time so that the right hand side of Eq.~\eqref{eq:irreversibility_bound} eventually becomes negative. For large times, the bound thus reduces to a trivial statement. For small times, the mean irreversibility can be calculated by taking the average of Eq.~\eqref{eq:irreversibility_smallTimes} which can be identified as the Kullback-Leibler divergence between the initial distribution and its $v$-reflected counterpart. In a NESS, where currents do not vanish, the distribution of $v$ for fixed $x$ becomes asymmetric. As a result, the Kullback-Leibler divergence can grow beyond any value when the driving increases and the right hand side becomes negative, which also renders this bound trivial. Overall, no universal insight can be gained from the irreversibility bound, Eq.~\eqref{eq:irreversibility_bound}.

\section{Free diffusion as a paradigmatic example \label{sec:freeDiffExample}}
To further investigate the behavior of the uncertainty product $\mathcal{Q}$, Eq.~\eqref{eq:uncertaintyProduct}, we consider the arguably simplest underdamped model: free diffusion in one dimension with drift. This is diffusion without an external potential, driven by a constant non-conservative force. The dynamics is given by the one-dimensional Langevin equation \eqref{eq:underdampedLangevin} and a constant force $F(x) = \Fe$. We project the motion in $x$ on a ring with perimeter $2\pi$ to get a unique steady state. As before, the initial conditions are sampled from the steady state distribution
\begin{equation}\label{eq:freeDiff_sspdf}
	p^\mathrm{ss}(x, v) = \frac{1}{2\pi} \sqrt{\frac{m}{2\pi T}} \exp\left[ -\frac{m}{2T} \left(v - \frac{\Fe}{\gamma}\right)^2\right].
\end{equation}

For the remainder of this manuscript, we restrict ourselves to the class of observables
\begin{equation}\label{eq:observable}
	Y_n(\T; w(x)) \equiv Y(\T; w(x) v^n) = \int_0^\T w(x) v^n dt
\end{equation}
and call the exponent $n \in \mathbb{N}$ the $v$-order of the observable. The corresponding uncertainty product for a current $Y_n$ is defined as
\begin{equation}\label{eq:uncprod}
	\mathcal{Q}_n(\T; w(x)) \equiv \mathcal{Q}(\T; w(x)v^n).
\end{equation}
Since we want to discuss the dependence on the external force, we write $\mathcal{Q}_n^\Fe$ in the following with a superscript $\Fe$ that indicates this parametric dependence.

By choosing $w(x)=1$, all moments occurring in the uncertainty product $\mathcal{Q}_n^\Fe(\T; 1)$, Eq.~\eqref{eq:uncprod}, can be calculated analytically. The first moment of $Y_n$ is simply given by $\la Y_n(\T; 1)\ra = \la v^n\ra \T $ where the ensemble average can be evaluated using the steady state distribution \eqref{eq:freeDiff_sspdf}. 
The second moment is defined by Eqs.~\eqref{eq:ode_YY_gen} and \eqref{eq:ode_Ywv_gen}. After inserting the corresponding weight $w(x,v) = v^n$, we obtain the time evolution of the second moment as
\begin{align}
	\frac{d}{d\T}\la Y_n^2\ra &= 2 \la Y_n v^n\ra \label{eq:ode_YY_free}\\
	\frac{d}{d\T}\la Y_n v^j\ra &= \la v^{n+j}\ra - \frac{j}{m}\left(\gamma \la Y_n v^j\ra - F_\mathrm{ext} \la Y_n v^{j-1} \ra\right) \nonumber \\
	&+ j(j-1) \frac{T\gamma}{m^2} \la Y_n v^{j-2}\ra \;\;\; \text{for } 1 \leq j \leq n \label{eq:ode_Yvj_free}
\end{align}
where we dropped the arguments of $Y_n$.

\begin{figure}
	\includegraphics[scale=1]{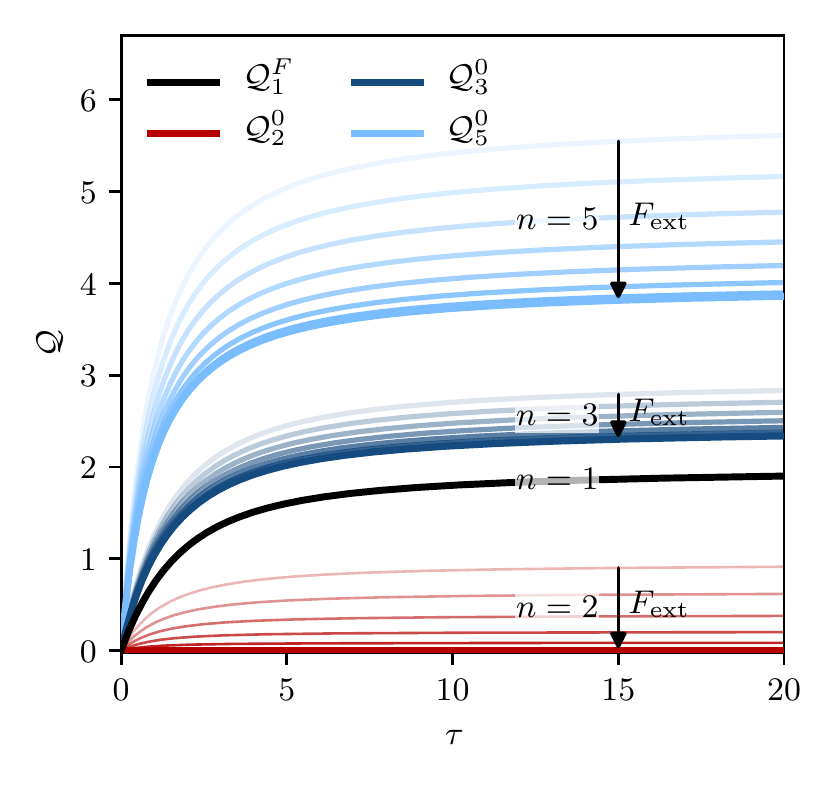}
	\caption{\label{fig:UP_comparison} Uncertainty product of free diffusion $\mathcal{Q}_n^\Fe(\T; 1)$ as defined in equation \eqref{eq:uncprod} for observables of the form $Y_n(\T; 1)$ (see Eq.~\eqref{eq:observable}) with $n \in \lbrace 1, 2, 3, 5\rbrace$ plotted against dimensionless time $\tau=\gamma\T/m$. The solid lines are the results for different forces $\Fe$ while the colors encode the $v$-order $n$. The thick black line is the exact result for $n=1$ which is independent from the applied force, see Eq.~\eqref{eq:uncProd_FreeDiff_n1}. The thick lines in the $n=3$ set and $n=5$ set correspond to the minimized uncertainty product, Eq.~\eqref{eq:uncProd_FreeDiff_n3} and  \eqref{eq:uncProd_FreeDiff_n5}, respectively. The thick line in the $n=2$ set corresponds to the limit $\Fe \rightarrow 0$ where $\mathcal{Q}_2^0$ is zero for all $\tau$.
	}
\end{figure}

This recurrent set of ordinary differential equations can be solved for any power $n$, beginning with the correlation $\la Y_n(\T; 1)v\ra$ with $j=1$. The corresponding integration constants are fixed by the condition that for $\T = 0$ all correlations $\langle Y_n v^j\rangle$ vanish.

We start the analysis with the lowest $v$-order $n=1$. In this case, the observable $Y_1(\T; 1)$ corresponds to the distance travelled in time $\T$. The solution of the uncertainty product for free diffusion according to Eqs.~\eqref{eq:ode_YY_free} and \eqref{eq:ode_Yvj_free} takes the form
\begin{align}\label{eq:uncProd_FreeDiff_n1}
	\mathcal{Q}^f_{1}(\T; 1) = \mathcal{Q}^0_{1}(\T; 1) &= \frac{2}{\tau}(\tau - 1 + \mathrm{exp}[-\tau] ) \\ &\approx 
	\begin{cases} 
      \tau & \tau\ll 1 \\
      2    & \tau \rightarrow \infty
    \end{cases} \nonumber
\end{align}
with dimensionless time $\tau \equiv \gamma \T/m$. Interestingly, the uncertainty product does not depend on the force $\Fe$.

As mentioned in Sec.~\ref{sec:underdampedDynamics}, the uncertainty product for free diffusion \eqref{eq:uncProd_FreeDiff_n1} is linear for small times which is a consequence of the ballistic evolution. For long times it asymptotically approaches $2$ from below. Consequently, the overdamped TUR is violated for all finite times. However, if we consider the overdamped limit of \eqref{eq:uncProd_FreeDiff_n1} where $\gamma/m \gg 1$ we indeed recover the expected behavior for overdamped free diffusion with the uncertainty product reaching $2$ for times $\tau \gg m/\gamma$.

For a higher $v$-order, the uncertainty product for free diffusion depends on the force $\Fe$ with a striking difference for odd and even observables. First, we discuss the odd case. As shown in Fig.~\ref{fig:UP_comparison}, $\mathcal{Q}^f_{3}$ decreases for $\Fe \rightarrow 0$ and eventually converges to a finite limit given by
\begin{align}\label{eq:uncProd_FreeDiff_n3}
	\mathcal{Q}^\Fe_{3}(\T; 1) \geq &\mathcal{Q}^0_{3}(\T; 1) \\
	&= \frac{2}{\tau}\left( \frac{11}{9}\tau - \frac{29}{27} + e^{-\tau} + \frac{2}{27}e^{-3\tau}\right) \nonumber
\end{align}
as indicated by the thick line in Fig.~\ref{fig:UP_comparison}. 

The minimum of the uncertainty product of free diffusion $\mathcal{Q}^0_{3}(\T; 1)$ with an observable of order $n=3$ has similar properties as the $n=1$ uncertainty product. In particular, it is linear for small times and converges to a finite long-time limit
\begin{equation}
	\mathcal{Q}^0_{3}(\T; 1) \approx 
	\begin{cases} 
      \frac{5}{3}\tau & \tau\ll 1 \\
      \frac{22}{9}    & \tau \rightarrow \infty .
    \end{cases}
\end{equation}
The steeper slope for small times and the larger value in the long-time limit compared to $\mathcal{Q}_1^0$ is due to the fact that the higher exponent in the weight increases the contribution of events in the vicinity of the typical value, thus increasing the variance of the observable without influencing its mean as strongly. This effect results in the uncertainty product increasing faster in the ballistic regime and in settling on a higher value in the long-time limit. Consistently, $\mathcal{Q}^0_{3}$ is larger than $\mathcal{Q}^0_{1}$ for all times.

The qualitative observations made for the observable scaling with $v^3$ are valid for the observable $Y_5(\T; 1)$ with $n=5$ as well. The bright lines in Fig.~\ref{fig:UP_comparison} show the uncertainty product over the dimensionless time $\tau$ for different forces. Again, a minimum is obtained in the equilibrium limit
\begin{align}\label{eq:uncProd_FreeDiff_n5}
	\mathcal{Q}^0_{5}(\T;1) &= \frac{2}{\tau}\left( \frac{449}{225}\tau - \frac{4447}{3375} + e^{-\tau} + \frac{8}{27}e^{-3\tau} + \frac{8}{375}e^{-5\tau}\right) \nonumber \\
	&\approx 
	\begin{cases} 
      \frac{21}{5}\tau & \tau\ll 1 \\
      \frac{898}{255}    & \tau \rightarrow \infty .
    \end{cases}
\end{align}
This result for $n=5$ lies above the uncertainty product for the lower $v$-orders $\mathcal{Q}_3^0$ and $\mathcal{Q}_1^0$.

For an even exponent $n$ in the weight the results look quite different, as plotted exemplarily for $n=2$ in Fig.~\ref{fig:UP_comparison}. In contrast to the previously analyzed odd exponents, the minimum of the uncertainty product in the limit $\Fe \rightarrow 0$ is $0$ for all times. The reason for this behaviour has already been discussed in Ref. \cite{VanVu:2019uu}. Since an observable with even exponent is even under time reversal, the time-reversed trajectory $(x(\T-t), -v(\T-t))$ of any realisation yields the same value for the observable $Y$. This generally leads to an non-vanishing mean value $\la Y\ra$ in the equilibrium limit. As a result, the uncertainty $\epsilon$ stays finite in this limit and the uncertainty product $\mathcal{Q}$ approaches $0$ when $\sigma$ becomes smaller. In contrast, for an odd observable, the time-reversed trajectory contributes the negative value and thus the original and time-reversed trajectory cancel in the calculation of the mean when they are equally probable. As a result, the uncertainty diverges in the equilibrium limit ultimately leading to a non-zero $\mathcal{Q}$.

The conceptual difference between time symmetric and antisymmetric observables is not unique to underdamped dynamics. For a Markovian jump dynamics on a discrete set of states the TUR holds only for odd, current-like observables. In this dynamics the precision of even observables, dubbed ``traffic'' or ``frenesi'', is not bounded by the entropy production but by the so-called time-symmetric dynamical activity \cite{garrahan_simple_2017}. Our results for free diffusion suggest that the same distinction is necessary for underdamped dynamics as well. While the uncertainty product of the considered odd, current-like observables is finite in the equilibrium limit and thus could be bounded by some non-trivial function, observables that are even under time-reversal approach $\mathcal{Q} = 0$ in this limit. A recent bound for underdamped Langevin dynamics which involves the dynamical activity \cite{Fischer:2018fi,VanVu:2019uu,Lee:2019uy} is valid for both even and odd observables but, unfortunately, is not very tight. In the following we will focus on bounds for current-like observables, searching for an underdamped generalization of the overdamped TUR.

\section{A conjecture based on free diffussion  \label{sec:freeDiffReference}}
As shown in Sec.~\ref{sec:freeDiffExample}, the uncertainty product of free diffusion becomes smaller than the original, overdamped formulation of the TUR, Eq.~\eqref{eq:overdamped_TUR} for all driving forces. In contrast to the overdamped TUR, the results obtained for free diffusion suggest that a putative bound should be time-dependent to cover the linear regime of the uncertainty product for small times. Consequently, the overdamped TUR cannot be straightforwardly generalized to underdamped motion. 

\subsection{The conjecture}
Conceptually, the original, overdamped TUR can also be interpreted as a bound generated by free diffusion. One of the features of the original TUR for overdamped dynamics is that it becomes saturated for free diffusion. In the proof the TUR follows from a bound on the large deviation function (LDF) \cite{Horowitz:2017ut} or on the scaled cumulant generating function \cite{Dechant:2018ga}. In both versions, the bounding function on the LDF (the generating function) is the one from free diffusion. In this sense, one could also say that the TUR states that the uncertainty product $\mathcal{Q}$ is bounded from below by the uncertainty product of free diffusion. The same interpretation holds for an approach to the TUR that is based on a Martingale decomposition~\cite{pigo17}.

In extensive numerical simulations for diffusion in a periodic potential we recognize the same relationship for underdamped motion. In detail, we find that the uncertainty product $\mathcal{Q}_n(\T; w(x))$ of an odd $n$-order current
\begin{equation}
	Y_n(\T; w(x)) = Y(\T; w(x)v^n), \;\; n\in \lbrace 1, 3, 5 ...\rbrace
\end{equation}
for a one-dimensional system described by the underdamped dynamics \eqref{eq:underdampedLangevin} without forces that depend on the velocity (e.g. the Lorentz force) is bounded from below by the respective result for one-dimensional free diffusion in the equilibrium limit with homogeneous weight $w(x) = 1$
\begin{equation}\label{eq:conjecture}
	\mathcal{Q}_n^{F(x)}(\T; w(x)) \geq \mathcal{Q}_n^0(\T; 1).
\end{equation}
We will refer to this conjecture as the \emph{free diffusion bound} (FDB). The right hand side of \eqref{eq:conjecture} has been calculated in Sec.~\ref{sec:freeDiffExample}. 

Most prominently, for the important class of currents $Y_1$ with $v$-order 1 the conjecture takes on the form
\begin{equation}\label{eq:conjecture_n1_current}
	\mathcal{Q}_n^{F(x)}(\T; w(x)) \geq 2 - \frac{2m}{\gamma\T}\left(1 - \mathrm{exp}[-\frac{\gamma}{m}\T] \right) .
\end{equation}

In the following we will describe the system in more detail and substantiate our conjecture by numerical data.

\subsection{Driven diffusion in a periodic potential \label{sec:perPot} }
We consider one-dimensional driven diffusion in a $2\pi$-periodic potential $V(x)$. The process is described by the Langevin equation \eqref{eq:underdampedLangevin} with scalar variables $x$ and $v$. The spatial coordinate $x$ is projected on a ring of perimeter $2\pi$ to get a unique steady state. The potential consists of sine and cosine modes up to second order and random amplitudes $c_i^\pm$ where the superscript $+$ ($-$) denotes the amplitude of the cosine (sinus) mode. In addition, a constant force $\Fe$ is applied. We randomly choose 500 parameter sets ($\Fe \in [0, 3.5]$, $T \in [0.5, 1.5]$, $\gamma \in [0.5, 5]$, $c_i^\pm \in [-2,2]$) and sample at least $50\,000$ trajectories of fixed length from the steady state for each set. The variance and mean value of the currents $Y_1(\T;1)$ and $Y_3(\T;1)$ are then computed for constant time along the different trajectories. Analogously, we extract the mean entropy production rate $\sigma$ by tracking the dissipation. 

\begin{figure}
	\includegraphics[scale=1]{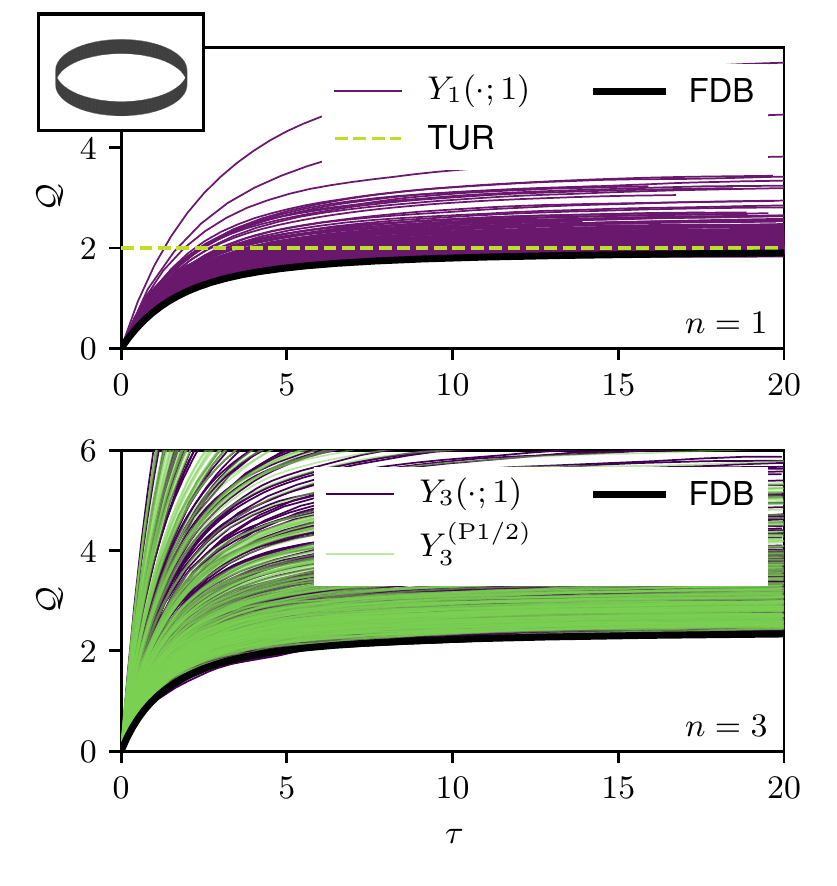}
	\caption{\label{fig:numerics_PerPot}Finite-time uncertainty product for different currents numerically evaluated for an underdamped particle on a ring. The upper panel shows the data for the current $Y_1(\T; 1) = \int v dt$ where each line corresponds to a diffusion process in a random potential characterized by its amplitudes $c_i^\pm \in [-2,2]$ and randomly sampled parameters with $\Fe \in [0, 3.5]$, $T \in [0.5, 1.5]$, $\gamma \in [0.5, 5]$. The conjectured bound, Eq.~\eqref{eq:conjecture}, is plotted as solid black line, the (overdamped) TUR and coincident the asymptotic behavior is indicated by the dashed line. The lower panel shows the uncertainty product for the current $Y_3(\T; 1) = \int v^3 dt$ and two exemplary $x$-dependent currents (P1) and (P2) (see Eqs.~\eqref{eq:perpot_3_coscurrent_1} and \eqref{eq:perpot_3_coscurrent_2}).}
\end{figure}

The simulation results for $Y_1$ are summarized in the upper panel of Fig.~\ref{fig:numerics_PerPot}, where each thin line corresponds to one parameter set. The result for free diffusion $\mathcal{Q}_1^0$, Eq.~\eqref{eq:uncProd_FreeDiff_n1} is plotted as thick, black line. Within the considered parameter range we see no violation of our conjecture, Eq.~\eqref{eq:conjecture}.

For $n=3$ we can establish the same role of free diffusion. In detail, the equilibrium limit of the uncertainty product obtained for free diffusion  $\mathcal{Q}_3^0$, see Eq.~\eqref{eq:uncProd_FreeDiff_n3}, bounds the uncertainty product from below for all times. We, again, validate this by randomly selecting 270 different parameter sets and plotting them as dark lines in the lower panel of Fig.~\ref{fig:numerics_PerPot}. To check that the conjecture holds with $x$-dependent weights as well, we furthermore evaluate the uncertainty product of the two currents
\begin{equation}\label{eq:perpot_3_coscurrent_1}
	Y_3^\text{(P1)}(\T) \equiv Y_3\left(\T;  1 + \frac{1}{2}\cos\left(2\pi x(t)\right) \right)
\end{equation}
and
\begin{equation}\label{eq:perpot_3_coscurrent_2}
	Y_3^\text{(P2)}(\T) \equiv Y_3\left(\T;  \cos\left(2\pi x(t)\right)^2 \right)
\end{equation}
which are plotted as bright lines in the lower panel of Fig.~\ref{fig:numerics_PerPot}.

\section{Case studies in higher dimensions}
So far, we have focused on one spatial dimension. In higher dimensions, it is not obvious  how to generalize our conjecture, Eq.~\eqref{eq:conjecture}, as different velocities of different directions can arise. The $v$-order of an observable involving different spatial dimensions is ambiguous as it can refer to either the overall order of all velocities or that of just one specific direction. In the following we will exemplarily study two different systems to examine the applicability of the FDB to higher dimensions. We emphasize that the results presented in the following are not conclusive, yet, but are rather intended as a starting point for further studies. 

\subsection{Underdamped diffusion on an torus}
First, we consider driven diffusion in a two-dimensional periodic potential, i.e. diffusion on a two-dimensional torus. The process is described by coupled Langevin equations for the variables $x_{1,2}$ and $v_{1,2}$ with periodic boundaries along both spatial dimensions. We apply the non-conservative force
\begin{equation}\label{eq:torus_force}
	\bm{F}(\bm{x}) = \left( c_1 \sin(x_1+x_2), c_2 \cos(x_1-x_2) \right)^T + \Febf
\end{equation}
with parameters $c_{1,2}$ and external driving $\Febf$.

The time-integrated current can, in principle, depend on all velocity components. First, we restrict the current to the projected velocity in either the first or the second direction
\begin{equation}\label{eq:torus_currents}
	Y_n^\text{(T1)}(\T) \equiv Y(\T; v_1^n) \;\; \text{and} \;\; 	Y_n^\text{(T2)}(\T) \equiv Y(\T; v_2^n)
\end{equation}
which correspond to a $v$-order $n=1$ and $n=3$, respectively. Furthermore, we consider the diagonal current
\begin{equation}\label{eq:torus_currents_T3}
	Y_n^\text{(T3)} \equiv Y_n^\text{(T1)} + Y_n^\text{(T2)}.
\end{equation}
As either term can dominate the sum, a lower bound that is based on the $v$-order, if existent, must be given by the smallest bound of the respective terms. This smallest bound corresponds to the term with the lowest occurring $v$-order. In this case, both terms in the current $Y_n^\text{(T3)}$ have the same $v$-order so that we attribute the $v$-order of $n$ to the observable (T3).

\begin{figure}
	\includegraphics[scale=1]{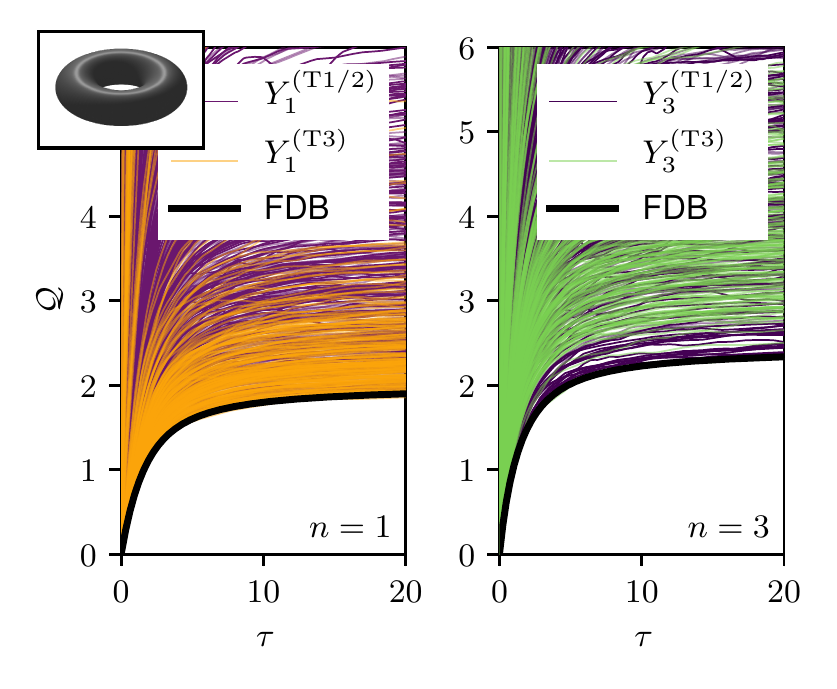}
	\caption{\label{fig:numerics_torus}The uncertainty product $\mathcal{Q}$ plotted against dimensionless time $\tau$ for diffusion on a torus. Each line corresponds to one of 250 parameter sets for $\gamma$, $T$ and the free parameters in the force \eqref{eq:torus_force}. The dark lines correspond to the projected currents in one direction $Y_n^\mathrm{(T1)}(\T)$, $Y_n^\mathrm{(T2)}(\T)$ (see Eq.~\eqref{eq:torus_currents}) while the bright lines give the diagonal current $Y_n^\mathrm{(T3)}(\T)$, Eq.~\eqref{eq:torus_currents_T3},  with $n=1$ in the left panel and $n=3$ in the right panel. The thick black line depicts the respective free diffusion bound $\mathcal{Q}_n^0$ that has been conjectured to be a lower bound for diffusion in one dimension.}
\end{figure}

We extract the uncertainty product as described in Sec.~\ref{sec:perPot}. The results of the three currents (T1) -- (T3) for $n=1$ and $3$ are shown in the both panels of Fig.~\ref{fig:numerics_torus}. The numerical data give a first indication that the respective free diffusion bound for one dimension, plotted as thick black lines, could be generalized to higher dimensions as well. 

The apparent validity of the conjecture is surprising as there is no straightforward mapping of the two-dimensional diffusion to a one-dimensional problem. Although the Langevin equation decouples for $c_1 = c_2 = 0$ and one arrives at effectively two one-dimensional processes, the diffusion is a genuine two-dimensional process in general. One could, however, argue that the additional degrees of freedom increase the uncertainty product. First, there might be dissipation due to directed motion in a direction that does not contribute to the considered current, which increases $\mathcal{Q}$. For instance for the current (T1) the force in the $2$-direction contributes only indirectly to the motion in $1$-direction while it directly increases the entropy production rate $\sigma$. Second, the potential mediates energy transfer between the two directions, thus increasing the fluctuations and also the uncertainty for the current in one specific direction. 

In principle, one can also consider mixed currents where both velocities are connected multiplicatively as in
\begin{equation}
	Y\left(\T; w(x) v_1^n v_2^m\right) = \int_0^\T w(x) v_1^n v_2^m dt\;\; n, m\in \mathbb{N}
\end{equation}
where one number $n$ or $m$ must be even and the other one odd in order to maintain the odd character under time-reversal. One example is the empirical correlation of the kinetic energy in $1$-direction and the velocity in $2$-direction measured along a trajectory. This quantity can be written in the form $Y(\T; m v_1^2 v_2 /2)/\T$. In contrast to the previously analyzed currents, it is not obvious how to define the ``$v$-order'' of such a current. On the one hand, one could argue that the odd part $v_1$ does only appear in first order. On the other hand, the overall exponent of velocities is $3$ thus suggesting that the uncertainty product can be estimated better by comparing with a one-dimensional $n=3$ current. We briefly address this issue exemplarily for the above current.

For a flat potential $c_i=0$ the uncertainty product can be solved analytically using the previously derived differential equations \eqref{eq:ode_YY_gen} and \eqref{eq:ode_Ywv_gen}. The corresponding uncertainty product for arbitrary driving $\Febf$, as before, depends on the force. In the equilibrium limit $\Febf \rightarrow \bm{0}$ the uncertainty product takes on the form
\begin{equation}\label{eq:underdampedTorus_mixedCurrent}
	\mathcal{Q}^{\bm{0}}\left(\T; \frac{m}{2} v_1^2 v_2\right) = \frac{2}{\tau}\left( \frac{5}{3}\tau - \frac{11}{9} + e^{-\tau} + \frac{2}{9}e^{-3\tau} \right)
\end{equation}
which is larger than both, the uncertainty product obtained for free diffusion of observables of order $1$ and $3$ in the equilibrium limit
\begin{equation}\label{eq:underdampedTorus_mixedCurrent_order}
	\mathcal{Q}^{\bm{0}}\left(\T; \frac{m}{2} v_1^2 v_2\right) \geq \mathcal{Q}_3^0(\T; 1) \geq \mathcal{Q}_1^0(\T; 1).
\end{equation}
The tighter bound for the $n=3$ current can be interpreted by considering the correlations between the velocities. Even when a potential mediates a correlation of the velocities in different spatial directions, the uncertainty of the current is still higher than that of a one-dimensional process where only one velocity exists. 

To see whether a bound holds in presence of a potential as well, we repeat the numerical analysis for the time-integrated current $Y(\T; m v_1^2 v_2 /2)$ and evaluate 200 random parameter sets numerically. All results lie above the value obtained without an external potential in the equilibrium limit, $\mathcal{Q}^{\bm{0}}\left(\T; m v_1^2 v_2/2\right)$. To increase the transfer of energy between the two spatial directions via the potential we further consider the conservative potential
\begin{equation}
	V(\bm{x}) = c_1 \sin(x_1 - x_2)
\end{equation}
which essentially forms a well of lower energy diagonally along the torus. Here, driving in the direction $x_1$ results in a consistent motion in the direction $2$ and vice versa. We simulate 100 different parameter sets with random values $c_1$, $\Febf$, $\gamma$, $T$, never observing an uncertainty product that goes below Eq.~\eqref{eq:underdampedTorus_mixedCurrent}. This finding suggests that a lower bound based on free diffusion holds for such mixed currents as well.

\subsection{The underdamped Brownian gyrator}
A model that is conceptually different from diffusion on a torus is that of a Brownian gyrator. It describes a charged particle in a two-dimensional harmonic potential with spring constant $k$ that is driven with a constant torque $\kappa$. The particle is embedded in a medium with friction coefficient $\gamma$ and temperature $T$. Furthermore, the particle is subject to the Lorentz force in presence of a magnetic field of strength $B$. Overall, the motion is described by the two-dimensional Langevin equation
\begin{align}
	\label{eq:underdampedLangevin_BrownGyr}
	\dot{\bm{x}}  &= \bm{v} \nonumber\\ 
	m\dot{\bm{v}} &= \left(\begin{matrix} -k & \kappa \\ -\kappa & -k \end{matrix}\right) \bm{x} + \left(\begin{matrix} -\gamma & B \\ -B & -\gamma \end{matrix}\right)\bm{v} + \bm{\xi}
\end{align}
with the usual statistics for the noise $\bm{\xi}$ (see Eq.~\eqref{eq:underdampedLangevin}). 

The stationary state of the linear Langevin dynamics \eqref{eq:underdampedLangevin_BrownGyr} can be solved exactly. The covariance matrix $\mathsf{C} = \la (\bm{x}, \bm{v})^\mathrm{T} (\bm{x}, \bm{v}) \ra - \la(\bm{x}, \bm{v})\ra^2$ is given by
\begin{equation}\label{eq:gyr_covariance}
	\mathsf{C} = \frac{T}{m \phi - \kappa^2 m/\gamma}
	\left(\begin{matrix} 
		\gamma 	& 0 		& 0 						& -\kappa \\
		0 		& \gamma	& \kappa					& 0 \\
		0		& \kappa	& \phi	& 0 \\
		-\kappa	& 0 		& 0							& \phi
	\end{matrix}\right)
\end{equation}
with $\phi \equiv (\gamma k + \kappa B)/m$. 

A natural current arising in this system is the observable
\begin{equation}\label{eq:gyr_obs1_Current}
	Y^\text{(G1)}_1(\T)
	\equiv \int_0^\T dt  \left[ x_2(t) v_1(t) - x_1(t) v_2(t) \right].
\end{equation}
which corresponds to the distance travelled in the gyrator. This circular current is also proportional to the work performed by the torque $\kappa$. Since the velocity appears in first order, the circular current can be regarded as an $n=1$ observable. Using the covariance matrix \eqref{eq:gyr_covariance} the mean value is given by
\begin{equation}\label{eq:gyr_n1_mean}
	\la Y^\text{(G1)}_1(\T) \ra = \T \frac{2\kappa T\gamma}{m( \gamma\phi - \kappa^2)}
\end{equation}
and using Eq.~\eqref{eq:entropyProduction} the entropy production rate can be expressed as
\begin{equation}\label{eq:gyr_entropy}
	\sigma = \frac{\kappa}{T} \partial_\T \la Y^\text{Gyr}_1(\T)\ra.
\end{equation}

For small times the variance can be calculated with Eq. \eqref{eq:VarY_smallTime}
\begin{align}
	\rm{Var}[&Y^\text{(G1)}_1(\T)] \approx \T^2 {\rm Var}[x_2 v_1 - x_1 v_2]  \\
		& = \T^2\left( \la x_2^2\ra\la v_1^2\ra + \la x_2v_1\ra^2 + \la x_1^2\ra\la v_2^2\ra + \la x_1v_2\ra^2\right). \nonumber
\end{align}
Here, the second line follows from Wick's theorem and from $\la x_1x_2\ra = \la x_iv_i\ra = 0$. Plugging the covariances in and identifying the mean current, Eq.~\eqref{eq:gyr_n1_mean}, finally yields
\begin{equation}\label{eq:gyr_n1_var}
	{\rm Var}[Y^\text{(G1)}_1(\T)] = \T^2 \la Y^\text{(G1)}_1 \ra^2 \frac{\gamma \phi + \kappa^2}{2 \kappa^2} + \mathcal{O}(\T^2).
\end{equation}

Combining the cumulants, we can express the uncertainty product in first order in time as
\begin{equation}\label{eq:gyr_uncProduct}
	\mathcal{Q}^\text{(G1)}_1(\T) = \T \frac{\gamma}{m} \frac{\gamma\phi + \kappa^2}{\gamma\phi - \kappa^2} + \mathcal{O}(\T^2) \geq \T\frac{\gamma}{m} + \mathcal{O}(\T^2)
\end{equation}
where the bound follows from minimizing in $\kappa$. More accurately the minimum is reached in the equilibrium limit $\kappa \rightarrow 0$ for arbitrary magnetic field. 

This short-time expansion coincides with the corresponding free diffusion bound \eqref{eq:uncProd_FreeDiff_n1} to first order in time. As a result, the conjectured bound \eqref{eq:conjecture} holds in the Brownian gyrator in this order with and without a magnetic field. It is due to the ballistic dynamics for small times, while the effect of the magnetic field sets in on a larger timescale.

\begin{figure}
	\includegraphics[scale=1]{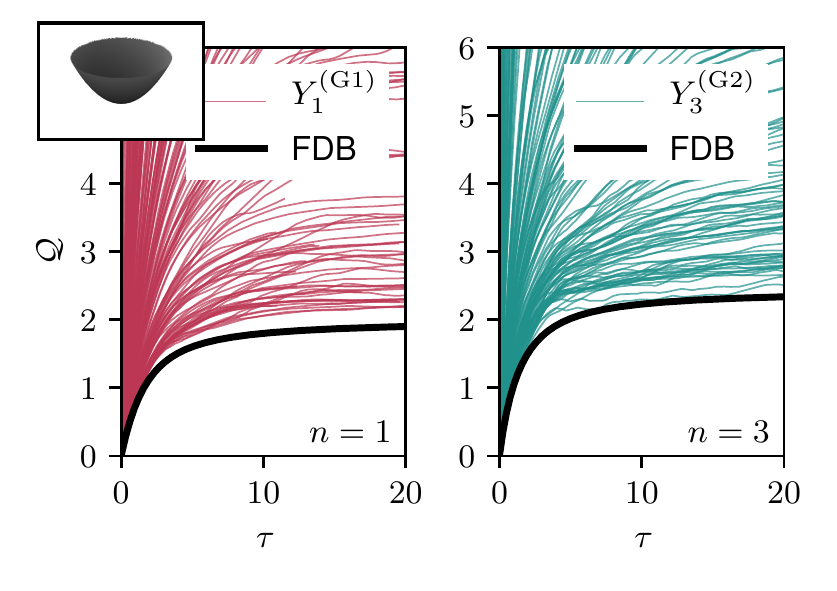}
	\caption{\label{fig:numerics_Gyr_Q} The uncertainty product for the two considered currents in the gyrator model, $Y_1^\text{(G1)}$ in the left panel and $Y_3^\text{(G2)}$ in the right panel, plotted against dimensionless time $\tau$. Each line corresponds to a different parameter set that was chosen on random in the absence of a magnetic field. The thick black line depicts the respective free diffusion bound $\mathcal{Q}_n^0$ that has been conjectured to be a lower bound for diffusion in one spatial dimension.}
\end{figure}

In presence of a magnetic field it is known that the uncertainty product in the long-time limit can become lower than $2$ and thus violates the conjectured bound for order $1$ currents. This effect has already been discussed in Ref.~\cite{Chun:2019hm}. In the following, we therefore consider the case without a magnetic field $B=0$ to investigate the uncertainty product and its relation to our conjecture in this system. We numerically compute the finite-time uncertainty product of $Y^\text{(G1)}_1$ for 160 randomly sampled parameters ($\gamma \in [0.5, 1.5]$, $T \in [0.5, 1.5]$, $k \in [0.05, 4]$ and $\kappa \in [0, 2.5]$) and plot them with respect to the dimensionless time $\tau$ in the left panel of Fig.~\ref{fig:numerics_Gyr_Q}. In agreement with our conjecture, all curves lie above the value obtained for one-dimensional free diffusion.

Motivated by the results for diffusion on a torus, we repeat the analysis for the more abstract order $n=3$ current of the form 
\begin{equation}\label{eq:gyr_obs3_Current}
	Y^\text{(G2)}_3(\T) \equiv \int_0^\T dt \left[ x_2(t) v_1(t)^3 - x_1(t) v_2(t)^3 \right].
\end{equation}
Using the same rationale as for the diagonal current on a torus, we consider this observable a current of $v$-order 3. The numerical results for 160 parameter sets with $\gamma \in [0.5, 5]$, $T \in [0.5, 1.5]$, $k \in [0.05, 3]$ and $\kappa \in [0, 2.5]$ are plotted in the right panel of Fig.~\ref{fig:numerics_Gyr_Q}. Again, in accordance with the conjecture the uncertainty product does not become smaller than the value obtained for free diffusion in one dimension in the equilibrium limit.

\section{Conclusion}
In this article we have addressed the issue of a thermodynamic uncertainty relation for underdamped dynamics. We analyze the finite-time uncertainty product for a Langevin process for the broad class of time-integrated observables consisting of the weighted mean of a function along the trajectory. This weight function can in general depend on both velocity and position.

By deriving the differential equation governing the time dependence of the second moment of the observable, we show that generally there exists a short-time, ballistic regime where the variance scales as second order in time. As a consequence, the uncertainty of a current is finite in the short-time limit and thus the uncertainty product is linear for small times. This linear behavior is quite different from the overdamped TUR, where the uncertainty product is larger than $2$ for all times.

To get an intuition for the typical time dependence of the uncertainty product, we have analyzed the arguably simplest model: free diffusion with drift in one dimension. We analytically find a qualitative difference between observables that are odd and observables that are even under time reversal. While the uncertainty product of odd observables reaches a finite value in the equilibrium limit, it becomes zero for even observables. This finding underlines that the distinction between odd current-like and even traffic-like observables that has been established for overdamped dynamics is relevant for underdamped dynamics as well. 

Based on numerical evidence we conjecture that the uncertainty product for an odd current with arbitrary weight function in a one-dimensional periodic potential is bounded from below by the result obtained for free diffusion for an observable of same order in the velocity but constant spatial weight in the limit of vanishing driving force. The conjectured bound converges to the overdamped TUR in the corresponding limit, thus suggesting that our conjecture is in fact the underdamped generalization of the TUR. By design, the bound is saturated for all times for free diffusion and thus is tight.

To our knowledge the conjectured free diffusion bound is the first bound that can be saturated for the important class of currents scaling with the first order in the velocity. Among this class of currents are, for instance, the integrated work current or the distance traveled in some time. Since such quantities can be measured experimentally, our bound can be used to infer bounds on the entropy production rate in systems where the latter is not directly accessible. In this context, it might be interesting to analyze for which current the conjecture becomes tightest as it was recently done for overdamped dynamics~\cite{busi19}, especially when the velocity dependence is taken into account. 

To assess how the conjecture can be generalized to higher dimensions, we have simulated two exemplary two-dimensional systems. We find that the results for free diffusion in one dimension bound the uncertainty product for these systems as well. The findings indicate, that it is possible to generalize the free diffusion bound to higher dimensions. However, further analysis will be necessary, especially regarding currents that contain velocities of different spatial directions multiplicatively. In this case, our data suggest that it is possible to get tighter bounds by adjusting the weight of the free diffusion process that is used for comparison.

With our bound being based on the result obtained for free diffusion, it might be possible to prove the conjecture by modifying the existing proofs for the TUR. These proofs are based on making suitable ansatzes for the arguments of the large deviation functional or for forces in a virtual dynamics. Adapting those to the behavior of free diffusion might lead to a proof of our conjecture. Finally, the link between free diffusion and the TUR, discussed here, might also help to better understand and further tighten existing (overdamped) bounds~\cite{Polettini:2016hu}.


\subsection*{Acknowledgement}
We thank Patrick Pietzonka for valuable discussions. H.-M. Chun has been supported by the National Research Foundation of Korea (Grant No. 2018R1A6A3A03010776).

\bibliographystyle{apsrev}

\end{document}